\documentclass[aps,prd,preprintnumbers,showpacs]{revtex4}
\usepackage[dvips]{graphicx}
\begin{document}

%
%

\eprint{Nisho-4-2017}
\title{Axion Stars and Repeating Fast Radio Bursts with Finite Bandwidths}
\author{Aiichi Iwazaki}
\affiliation{International Economics and Politics, Nishogakusha University,\\ 
6-16 3-bantyo Chiyoda Tokyo 102-8336, Japan.}   
\date{July. 15, 2017}
\begin{abstract}
We have proposed a model of non repeating fast radio bursts ( FRBs ); 
the collisions between axion stars and neutron stars
generate the bursts.  In this paper,
we propose a model of repeating FRBs which shows that they arise from the 
several collisions between magnetized accretion disk of a black hole and an axion star orbiting the black hole.
There would be many axions stars condensing as dark matter in an early stage galaxy so that such collisions arises
repeatedly. 
The radiations are emitted by coherent oscillations of electrons in the accretion disk. The oscillations are caused by 
oscillating electric fields, which are induced by axion stars 
under strong magnetic field $\sim O(10^{11})$G. 
The emissions are terminated by the thermal fluctuations of the electrons which result from
the thermalization of the oscillation energies. Although the radiations are 
monochromatic, the thermal Doppler effects broaden the spectral lines; 
their spectra $S(\nu)$ are given by $S(\nu) \propto \exp(-(\nu-\nu_c)^2/2(\delta\nu)^2)$
with the center frequency $\nu_c$.
The bandwidth $\delta\nu=\nu_c\sqrt{T_c/m_e}$ with electron mass $m_e$ is determined by the critical temperature $T_c$ 
at which the thermal fluctuations terminate the coherent emissions. 
The observed bandwidths $(3\sim5)\times10^2$MHz are originated from the strong magnetic fields $\sim O(10^{11})$G.
With such strong magnetic fields, large amount of the burst energies are produced.
Various center frequencies $\nu_c$ of the bursts in the repeating FRB 121102 come from the various rotation speeds in 
the disk which make an intrinsic frequency of the bursts being Doppler shifted.
The recent observation showing an
extreme magneto-ionic environment around the source of FRB 121102 supports our model.
\end{abstract}
\hspace*{0.3cm}

\hspace*{1cm}

\maketitle


\section{introduction}
Since a mysterious radio burst has been originally reported\cite{frb} in 2007,
more than $30$ fast radio bursts ( FRBs ) have been observed.
The durations of the bursts are typically
a few milliseconds. The sources of the bursts have been suggested to be
extra-galactic because of their large dispersion measures.
This suggests that the large amount of the energies $\sim 10^{40}$erg/s
are produced at the radio frequencies.
The event rate of the burst is roughly estimated to be $\sim 10^{-3}$ per year in a galaxy.
Repeating emissions from single source were not observed. 
They are called as non repeating FRBs.
But, the repeating emissions from a single source FRB 121102 have recently been discovered\cite{frb2,rep,2GHz} and
the follow up observations\cite{re3G,re1.6G,op} have shown that the bursts are originated 
in a dwarf galaxy with redshift $z\simeq 0.19$. The galaxy is conjectured to be an AGN.
The FRB has been observed 
with three bands $1.2$GHz $\sim 1.5$GHz, $1.7$GHz $\sim 2.3$GHz and $2.5$GHz $\sim 3.5$GHz.

The peculiar feature is that the bursts shows spectral volatility\cite{2GHz}.
Some of them show positive spectral indexes $ \alpha>0 $, while some show negative indexes $\alpha<0$
with their fluxes
$S_{\nu} \propto \nu^{\alpha}$ 
in frequency $\nu$. It seems that  
the feature is common to non repeating FRBs.
Furthermore, the recent observations \cite{finite} with multi telescopes show
that spectra of the repeating bursts are not modeled by
a power law $\nu^{\alpha}$, but  Gaussian $\exp(-(\nu-\nu_c)^2/2(\delta\nu)^2)$.
The Gaussian forms of the spectra must lead to the spectral volatility and
suggest an intrinsic origin of the emission mechanism of FRBs. That is,  
intrinsic radiations are monochromatic but thermal Doppler effects broaden their spectral lines.
( The recent report of a non repeating extremely bright FRB 170107 \cite{170107}  shows sharp cutoff above $1400$MHz.
The cutoff is speculated to come from the Gaussian form of the spectrum. ) 

Up to now, there are many models proposed\cite{model} for the production mechanism of FRBs. 
Some of them are cataclysmic such as neutron star-neutron star mergers,
while some are not cataclysmic such as flare of young magnetars.
It is generally believed that the presence of the repeating FRB rules out
the cataclysmic models. Then,
our previous model\cite{iwazaki,t} in which FRBs arise from the collisions between axion stars and neutron stars 
can not explain the repeating FRB 121102.
Up to now,  there are still not promising astrophysical models for FRBs.

\vspace{0.1cm}
In this paper we propose a model for a production mechanism of the repeating FRB, assuming the presence of QCD axions as a dark matter. The FRB arises from
the collisions between strongly magnetized accretion disk of a black hole and axion star orbiting the black hole.
We suppose that most of the dark matter axions form axion stars, some of which 
condense in a center of early stage galaxy where a black hole is present. 
Orbiting the black hole, they
collide the magnetized accretion disk several times and finally evaporate or are absorbed in the black hole.
The bursts arise each time the collision takes place.
Using the model, we can naturally understand the Gaussian spectra with narrow bandwidths and various center frequencies of
the repeating bursts as well as large amount of radiation energies. 

\vspace{0.1cm}
Here we assume that the accretion disk is geometrically sufficiently thin in order for the axion star to be able to pass the disk. 
Such a disk would have much low temperature $T$. 
We also assume that 
magnetic field involved in the disk is very strong such as those of neutron stars e.g. $10^{11}$G.
Actually, there is a model\cite{accretion} of geometrically thin accretion disk with strong magnetic fields $B\sim 10^{11}$G 
at the site near black holes whose masses $M=10^2M_{\odot}\sim 10^4M_{\odot}$. In the model the thickness of the disk is
approximately given by $10^5\mbox{cm}(M/10^3M_{\odot})(T/10^2\rm eV)^{1/2}$ at the site near the black holes.
Therefore, we expect that the disk may be sufficiently thin for the axion star to be able to pass through the disk without
evaporation.
Tentatively, we adopt these parameters in the discussions below.  

Since the accretion disk rotates around a black hole with high velocities, the radiations emitted by the disk are Doppler shifted ( blue shifted 
in the case of the FRB 121102).
As a consequence, the FRB 121102 is observed with various frequencies $1.2\rm GHz\sim 3.5Ghz$,
depending on the velocities with which the disks at impact sites rotate. 
( We assume that 
the intrinsic one $\nu_{\rm int}$ may be in a range $1.2$GHz to $1.4$GHz. )

\vspace{0.1cm}
The essence of our model is that
axions\cite{axion} are converted into radiations under strong magnetic field. 
Thus, the coherent states ( e.g. axion stars ) of the axions can produce coherent radiations like FRBs through electrons.
The radiations are monochromatic with the frequency $\nu_{\rm in}$ given by the axion mass $m_a$; $\nu_{\rm in}=m_a/2\pi$.
They are emitted by coherently oscillating electrons in the disk. 
We show that the thermalization of the oscillation energies of the electrons terminates 
the coherent radiations possibly within a millisecond. As a result,
the line spectrum is thermally broadened.
The repeating bursts arise in the accretion disk with magnetized electron gases around a black hole.
The model for the repeating FRB 121102 is supported by the recent observation\cite{rotation}
which shows the presence of an extreme magneto-ionic environment around the source of FRB 121102.

Hereafter we use the unit of light velocity $c=1$ and the Boltzmann constant $k=1$.

\vspace{0.1cm}
In the next section (\ref{2}), we briefly explain axions and axion stars. We explain our model for a production mechanism of FRBs in the section (\ref{3}).
We show that when the axion stars collide magnetized electron gases, they induce coherent oscillations of the electrons which emit
coherent radiations, that is, FRBs.  We also show that because temperatures of the electrons rapidly increase, 
the coherent oscillations are disturbed by thermal fluctuations so that
the coherent emissions are terminated within a millisecond.
In the section ( \ref{4} ) we discuss the conditions for the axion stars to pass through the accretion disks without losing their whole energies. 
We show that the large amount of energies of FRBs are produced when the axion stars collide accretion disks with
electron gases under strong magnetic fields $\sim 10^{11}$G.
In the section (\ref{5}),
we explain several features of the FRB 121102. Especially, we explain that the radiations show
thermally broadened spectra $\sim \exp(-(\nu-\nu_c)^2/2(\delta\nu)^2)$ with $\delta\nu\propto \nu_c$.
We also show that their bandwidths $\delta\nu=300\rm HMz\sim 500$HMz for $\nu_c=2\mbox{GHz}\sim 3\mbox{GHz}$ originate in
the accretion disk with strong magnetic fields $\sim 10^{11}$G. 
In the final section (\ref{6}) we discuss the relation between the observed frequencies $\nu_{\rm ob}$ and
the intrinsic frequency $\nu_{\rm int}$, and summarize our results. 

\section{axion}
\label{2}
The axion is a Nambu-Goldstone boson\cite{axion} associated with Pecci-Quinn symmetry. 
The symmetry naturally solves the strong CP problem in QCD.
Although the axion is the Nambu-Goldstone boson, it acquires 
mass $m_a$ through chiral anomaly because the Pecci-Quinn symmetry is chiral. 
Thus, instantons in QCD give rise to the mass of the axion. 

The axions are one of most promising candidates of
dark matter. 
The axions may form axion stars
known as oscillaton\cite{axion2,osci,bs} made of axions bounded gravitationally.
In the early Universe,
axion miniclusters\cite{kolb}
are produced just after the QCD phase transition and grow in matter dominated period.
They may be the dominant component of dark matter in the Universe.
Furthermore, by losing their kinetic energies they may form 
gravitational bound states of the axions. Namely, 
the axion stars are formed as further condensed objects\cite{kolb,axions,iwa} 
of the axion miniclusters. 
We have recently proposed\cite{iwazaki,iwazaki2} a model for the production mechanism of the non repeating FRBs;
the FRBs arise from the collisions between the axion stars and neutron stars.  
Almost all of the properties ( event rate, large amount of radiation energies and absence of bursts with much high or low frequencies
like x ray or radio waves with frequencies $<100$MHz ) can be naturally explained in our model.
But, this mechanism can not explain the presence of the repeating FRB 121102.

\vspace{0.1cm}
As the axion field is real scalar, 
there are no static solutions representing the axion star in a system of the axion coupled with gravity. 
( i.e. solutions in the coupled equations of axion field equation and Einstein equations. )   
But, there are oscillating solutions\cite{axion2,osci,bs} of the axion stars.
They are coherent states representing gravitational bound states of the axions.
When the gravitational binding energies of the axions are much smaller than the axion mass,  
the approximate solutions\cite{iwazaki2} representing spherical axion stars with mass $M_a$ are given by 

\begin{equation}
\label{a}
a=a_0f_a\exp(-\frac{r}{R_a})\cos(m_a t) \quad \mbox{with} 
\quad a_0=1.2\times 10^{-7}\Bigl(\frac{3.5\times10^2\mbox{km}}{R_a}\Bigr)^2\frac{0.6\times10^{-5}\mbox{eV}}{m_a}
\end{equation}
with the decay constant $f_a$ of the axions, 
where the radius $R_a$ of the axion stars is approximately given by
\begin{equation}
R_a=\frac{1}{GM_a m_a^2}\simeq 
360\mbox{km}\Bigr(\frac{0.6\times10^{-5}\mbox{eV}}{m_a}\Bigl)^2\frac{2\times10^{-12}M_{\odot}}{M_a}
\end{equation}
where $G$ denotes the gravitational constant.
The decay constant $f_a$ is related with the mass $m_a$; 
$m_a\simeq 6\times 10^{-6}\mbox{eV}\times (10^{12}\mbox{GeV}/f_a)$. 
( More precise solutions can be seen in the reference\cite{sol}. 
But, the precise forms,  especially radial dependences of the solutions are not necessary to our discussions below. )

The point needed in our discussions is the dependence of the value $a_0=a/f_a$ on $M_a$ ( or $R_a$ ) and $m_a$. 
The dependence of $a_0=a/f_a$ on $M_a$ and $m_a$ can be approximately determined 
by the formula $M_a=\int d^3x ((\partial_t a)^2+(\vec{\partial}_x a)^2+(m_aa)^2)/2 
\sim \int d^3x (m_aa)^2 \sim (m_a a)^2\times \mbox{(volume of axion star)}$. Obviously,
the formula holds for the field configurations $a$ with much low momenta $\partial_x\simeq k\ll m_a$ and
even with field configurations with arbitrary shapes.  
In the spherical shape of the solutions in eq(\ref{a}), we have the volume given by $4\pi R_a^3/3$ and have 
the relation $R_a=\frac{1}{GM_a m_a^2}$.
Thus, we obtain $a/f_a\propto 1/(R_a^2m_a)$. 
The dependence of $a/f_a$ on $M_a$ ( or $R_a$ ) and $m_a$ is used for estimation of radiation energies of FRBs. 
We also point out that the value $a\sim M_a^{1/2}(m_a^2\times \mbox{volume of axion star})^{-1/2}$ do not significantly change even if
the axion star is deformed or split off by tidal forces of neutron stars or black holes. 
This is because the tidal forces do not change the mass $M_a$ and also do not significantly change the volume of axion star:
The small change of the volume leads to negligible change of $a/f_a$ because $a/f_a\propto (\sqrt{\mbox{volume}})^{-1}$.
( The relation  $R_a=\frac{1}{GM_a m_a^2}$ can be approximately obtained in the following.
That is, the binding energy $k^2/2m_a$ of the axion is equal to $-GM_am_a/R_a<0$, where the imaginary momentum $k=i/R_a$
characterizes the wave functions $a(r)\sim \exp(i k r)=\exp(-r/R_a)$. It follows that $R_a \simeq 1/GM_a m_a^2$. )

\vspace{0.1cm}
There are two unknown parameters $m_a$ and $M_a$ ( or $R_a$ ). 
The mass $M_a$ of the axion star was obtained\cite{iwazaki} by comparing the rate of the collisions between neutron stars
and axion stars in a galaxy with the event rate of FRBs $\sim 10^{-3}$ per year in a galaxy.
The mass takes a value in the range given by $M_a=(10^{-11}\sim 10^{-12})M_{\odot}$.
We should notice that the event rate of the FRBs has still not been precisely determined.
It gives rise to an ambiguity in the determination of $M_a$. 
Similarly we do not know the precise value of the axion mass $m_a$.
We anticipate that the mass take a value in the range of $m_a=0.6\times10^{-5}\mbox{eV}\sim 1.0\times10^{-5}$eV. 
Here we tentatively suppose that
$m_a$ is given by $\simeq 0.6\times10^{-5}$eV in order for 
the intrinsic frequency $\nu_{\rm int}=m_a/2$ of non repeating FRBs to be equal to
$1.4$GHz. 
( The observed frequencies are the ones of radiations affected by thermal Doppler effects
and cosmological redshifts. )   
Hereafter,
these parameters ( $M_a=2\times 10^{-12}M_{\odot}$ and $m_a=0.6\times 10^{-5}$eV ) are tentatively used 
in our discussions, keeping the presence of large ambiguities in $M_a$ and $m_a$ in mind.

( Sometimes, boson stars\cite{bs} are addressed using complex scalar fields. But, there is a conserved quantity in models of the complex scalar fields,
that is,  the number of quantum represented by the fields. Thus, their features are quite different from those of
the axion stars described by the real scalar field, in which there are no such conserved quantities. Additionally, there are papers\cite{bs}
addressing with axion stars by using equation of complex scalar field as a component with positive or negative frequency of quantized real scalar field.
The axion stars are coherent states of the axions so that the expectation values of annihilation operators
are identical to those of creation operators. 
Thus, the relevant field equation describing the axion stars is an equation of real scalar field, not complex scalar field. 
Their dependence on time arises through functions such as $\sin(\omega t)$ or $\cos(\omega t)$, not $\exp(i\omega t)$. 
When the axion stars are described as eigenstates of axion number operator,
the expectation value of the axion field vanishes; $\langle a \rangle=0$. The field equation becomes trivial.) 

\vspace{0.1cm}
We should make a comment that the solution in eq(\ref{a}) represents a coherent state of the axions.
The number of the axions in a volume with $m_a^{-3}$ ( $m_a^{-1}$ is Compton wavelength of the axion ) is extremely large; $M_a/m_a(m_a^{-3}/R_a^3)\sim 10^{41}$ 
with $M_a=10^{-12}M_{\odot}$.
The coherence is very rigid. Then, when the axion star approaches neutron stars or black holes,
their tidal forces distort the shape of the axion star like long sticks\cite{iwazaki3}. But the coherence is kept owing to the huge number of the axions
in the volume $m_a^{-3}$. Because the tidal force does not change the mass of the axion stars, 
we may use the average value $\bar{a}$ given by the approximate formula 
$M_a=\int d^3x ((\partial_t a)^2+(\vec{\partial}_x a)^2+(m_a a)^2)/2\sim (m_a \bar{a})^2\times \mbox{( volume of the axion star )}$,
even when the spherical form in eq(\ref{a}) is deformed. 
The tidal forces may change the volume of the axion star. But even if the volume becomes $10$ times larger, the value $\bar{a}$ only changes to be 
$3$ times smaller.  Moreover,   
as we explained, there are large ambiguities of the value $M_a$.
Additionally,
axion stars may have different masses with each other.  
These facts lead to ambiguities in the value $\bar{a}$ used in the present paper.
Although there are these ambiguities,
our purpose is to show that we can 
understand almost of all features of the FRBs, using the parameters mentioned above.


\section{emission mechanism of FRB}
\label{3}
The point is that
the axion star generates an electric field $\vec{E}$ when it is put in an external magnetic field $\vec{B}$.
In order to show it, we note
that
the axion couples with both electric $\vec{E}$ and magnetic fields $\vec{B}$ in the following,

\begin{equation}
\label{L}
L_{aEB}=k_a\alpha \frac{a(\vec{x},t)\vec{E}\cdot\vec{B}}{f_a\pi}
\end{equation}
with the fine structure constant $\alpha\simeq 1/137$,   
where the numerical constant $k_a$ depends on axion models; typically it is of the order of one.
Hereafter we set $k_a=1$.
From the Lagrangian, we derive the Gauss law, 
$\vec{\partial}\cdot \vec{E}=-\alpha\vec{\partial}(a\vec{B})/f_a\pi+\mbox{charge density}$. We have an additional term associated with
the axion-electromagnetic interaction.
When the axion star described by $a$ is imposed under background magnetic field $\vec{B}$,
we find from the Gauss law that the electric field is generated on the axion star, 

\begin{eqnarray}
\label{ele}
e\vec{E}_a(r,t)&=&-\alpha \frac{a(\vec{x},t)e\vec{B}}{f_a\pi}
=-\alpha \frac{a_0\exp(-r/R_a)\cos(m_at)e\vec{B}(\vec{r})}{\pi}\\
&\simeq& 5.6\times 10^{-1}\,\mbox{eV}^2( \,\,\simeq2.8\times 10^4\mbox{eV}/\mbox{cm} \,\,)\cos(m_at)
\Big(\frac{3.5\times10^2\,\mbox{km}}{R_a}\Big)^2\frac{0.6\times10^{-5}\mbox{eV}}{m_a}\frac{eB}{10^{11}\mbox{G}}\frac{\vec{B}}{B}.
\end{eqnarray} 

The electric field $\vec{E}_a$ is parallel to the magnetic field $\vec{B}$ and
coherently oscillates over the region with the magnetic field. Thus,
when the axion stars are put in magnetized ionized gases, e.g. electron gases,
the electric field $\vec{E}_a$ induces harmonically oscillating electric currents which emit radiations.
The radiations are linearly polarized dipole radiations. 
They are coherent radiations so that  
the radiations with large amount of energies are emitted with the frequency $m_a/2\pi$.
( Each electron oscillates with the identical phase to those of other electrons. Thus, the coherent emissions
arise. ) Because the magnetic fields are supposed to be strong e.g. $10^{11}$G, the electric fields are sufficiently strong for the radiations to have
the large amount energies.  
This is our production mechanism of FRBs. The mechanism can be applied to both non repeating FRBs and the repeating FRB 121102.
The non repeating FRBs are emitted in atmospheres of neutron stars while bursts in the repeating FRB are
emitted in magnetized accretion disk around a black hole.
As we show, the bursts in our model are dipole radiations. Indeed, 
linearly polarized radiations have recently been clearly observed\cite{rotation} in the FRB 121102.

A comment is in order. It is well known that an axion with momentum $\vec{k}$ is transformed into an electromagnetic wave with momentum $\vec{k}$
under static magnetic field with large spatial extension $R\gg k^{-1}$. The coherent axions are
transformed into coherent electromagnetic fields. In the case of the axion stars, which are composed of coherent axions with small momentum ( $k\simeq 1/R_a \ll m_a$ ),
such electromagnetic field is just the oscillating electric field in eq(\ref{ele}). 
    
\vspace{0.1cm}
The magnetized ionized gases are present in atmospheres of neutron stars or in accretion disks around a black hole.  
The radiations are non repeating FRBs when the axion star collides with a neutron star, while they are
repeating FRBs when the axion star collides with a magnetized accretion disk:
The collisions may occur several times because the axion star orbits the black hole. 
On the other hand, the collisions with neutron stars occur only once because the axion stars are absorbed in the neutron stars
after the collisions.

We note that the motion of electrons accelerated by the electric field $\vec{E}_a$ 
are restricted in one dimensional direction parallel to the electric and magnetic field.
This is because the cyclotron energy is larger than the temperature $T$ of the electrons;  
$eB/m_e\simeq 4\times10^3\mbox{eV}(eB/10^{11}\mbox{G})>T=10^2\mbox{eV}(T/10^6\mbox{K})$.
( The equation of the motion in the direction is given by $dp/dt=eE$ so that the momentum of electron is 
given by $p\simeq eE/m_a\simeq eE(t=0)\cos(m_at)/m_a$. )
It is important to notice that 
the energies $p^2/2m_e\simeq (eE)^2/m_a^2m_e$ of the oscillating electrons caused by the electric field are partially absorbed in the electron gases. 
In other words, the energies are thermalized.
As a result, the temperatures of the electrons
rapidly grow with time. Eventually,
thermal fluctuations disturb and terminate the coherent one dimensional oscillations.
The critical temperature $T_c$ at which thermal fluctuations terminate coherent emissions is given by
$T_c \sim p^2/2m_e\sim (eE)^2/m_a^2m_e\sim 2\times 10^4\mbox{eV}(eB/10^{11}G)^2$. That is, the thermal energy $T_c$
is equal to the oscillation energy $p^2/2m_e$.
The duration of the coherent radiations might be within a period of the order of a millisecond or less.    

\vspace{0.1cm}
The duration depends on the way how the temperatures of the electron gases increase when the oscillation energies
are thermalized.
To obtain the precise values of the duration we need to know the natures ( e.g. heat capacity ) of the gases in detail. Here,
we would like to show that mean free time of electrons is much less than a millisecond. 
Thus, within a millisecond, the oscillating electrons interact with each other
and so their kinetic energies could be thermalized within a millisecond. 
This is a necessary condition for the coherent emission
to be terminated by the thermal fluctuations within the period.
But it suggests that the coherent emissions stop within a millisecond.

\vspace{0.1cm}
Mean free times of oscillating electrons are obtained in the following.
The collision cross section $\sigma=\pi r^2$ of the electrons
is determined by the equality of Coulomb energy and kinetic energy,
$\alpha/r^2\simeq p^2/2m_e$. That is, they collide when they approach each other within a distance $r=\sqrt{2m_e\alpha/p^2}$.
Because mean free path $l$ is given such that $l=1/n_e\sigma$, we obtain mean free time $\tau=l/(p/m_e)$,

\begin{equation}
\label{mf}
\tau=\frac{p^3}{4\pi \alpha^2 m_en_e}\simeq 0.2\times 10^{-5}\mbox{sec}\Bigl(\frac{eB}{10^{11}G}\Bigr)^3 \Bigl(\frac{10^{17} \mbox{cm}^{-3}}{n_e}\Bigr).
\end{equation}

The formula holds only when the kinetic energies of the oscillating electrons are much larger than the thermal energies $T$.
If the temperatures increase and approach the critical temperature $T_c$, the formula does not hold.

We find that the electrons interact with each other many times within a millisecond.
So we expect that within a millisecond, the oscillation energies are thermalized 
and the coherent emissions are terminated by the thermal fluctuations.

A comment is in order.
The mean free time $\tau$ should be longer than the oscillation period $2\pi/m_a\sim 10^{-9}$ sec and shorter than a millisecond. 
If $\tau$ is shorter than the oscillation period, the approximation used for the derivation of $\tau$ is not valid.
If it is longer than a millisecond, the thermalization of the oscillation energies does not arise within a millisecond.
It leads to the duration of the bursts longer than a millisecond.

\vspace{0.1cm}
We would like to mention that
the identical values of the parameters $eB$ or $n_e$ used for the estimation of the mean free time in eq(\ref{mf})  
are also used for
the estimation of radiation energies, narrow bandwidths of their spectra, et al..
We should stress the notable fact that the almost identical values of the parameters used in our model lead to 
known observed values of the FRBs, as are shown below.

\section{conditions for survival in collision with accretion disk}
\label{4}
Now, we wish to discuss conditions on which the axion stars can pass through the accretion disk when they collide the disk.
Namely, it is the condition for them not to lose their whole energies in the collision.

The coherent radiations arise from oscillating electrons in the volume $\lambda_a^3\simeq (20\rm cm)^3$. 
$\lambda_a$ denotes the wave length of the radiations; $\lambda_a=1/\nu_{\rm int}=2\pi/m_a$ with $m_a=0.6\times 10^{-5}$eV.
They are dipole radiations.
Thus, the energies $\dot{W}$ of the radiations emitted in unit time from the volume are given by

\begin{equation}
\label{W}
\dot{W} 
\sim 10^{33}\,\mbox{GeV/s}\,\Big(\frac{n_e}{10^{17}\mbox{cm}^{-3}}\Big)^2
\,\Big(\frac{3.5\times10^2\rm km}{R_a}\Big)^4\Big(\frac{0.6\times10^{-5}\rm eV}{m_a}\Big)^8\Big(\frac{eB}{10^{11}\mbox{G}}\Big)^2,
\end{equation}
where we used the radiation energy $e^2\dot{p}^2/(12\pi m_e^2)$ emitted in unit time from a harmonically oscillating single electron.
The emissions reduce energies of axion stars.
On the other hand, the axion energy involved in the volume $\lambda_a^3$ of the axion star is given by $M_a(\lambda_a^3/R_a^3)$. 
It takes a time $h/v_0$ for the axion star to pass the disk with thickness $h$; $v_0$ denotes the velocity of the axion star. 
That is, the emissions last for the period $h/v_0$.
Hence, the axion star can pass through the disk without evaporation 
if the condition,
$\dot{W}(h/v_0)<M_a(\lambda_a^3/R_a^3)$, is satisfied.
Numerically,  
\begin{eqnarray}
\label{para}
&&\dot{W}\frac{h}{v_0}\sim 10^{27}\,\mbox{GeV/s}\,\Big(\frac{n_e}{10^{17}\mbox{cm}^{-3}}\Big)^2
\,\Big(\frac{3.5\times10^2\rm km}{R_a}\Big)^4\Big(\frac{0.6\times10^{-5}\rm eV}{m_a}\Big)^8\Big(\frac{eB}{10^{11}\mbox{G}}\Big)^2
\Big(\frac{h}{10^4\mbox{cm}}\Big)\Big(\frac{10^{10}\mbox{cm/s}}{v_0}\Big) \nonumber \\
&&M_a\frac{\lambda_a^3}{R_a^3}\sim 2\times10^{-12}M_{\odot}\frac{\lambda_a^3}{R_a^3}\simeq 0.4\times 10^{27}\rm GeV.
\end{eqnarray}
We can find that if the magnetic field is weaker than $10^{11}$G or the electron density is lower than $10^{17}$cm$^{-3}$
( or the thickness $h$ is smaller than $10^4$cm), the axion star can pass through the disk without the evaporation.

If we take much small physical parameters, e.g. small $eB\sim 10^8$Gauss for the axion stars to survive the collision, 
the emission energies are 
not sufficiently large to explain the observed values. But, 
we can easily see that when we take the parameters for the axion star just like ones used in eq(\ref{para}) to survive the collision, 
the radiations with sufficiently large amount of energies $\dot{W}(R_a^2/\lambda_a^2)\sim 10^{42}$erg/s are produced.
The energies are roughly consistent with the observed energies $\sim 10^{40}$erg/s.
( The formula $\dot{W}R_a^2/\lambda_a^2=\dot{W}\lambda_a R_a^2/\lambda_a^3$ represents the emission rate from the surface with depth $\lambda_a$ of the accretion disk.
The only radiations from the surface can arrive at the earth. That is,
although the emission can arise even inside of the disk, 
the radiations are absorbed inside of the disk itself and do not arrive the earth. We should notice that  
the axion stars 
are stretched by the tidal forces of neutron stars or black holes just like long sticks\cite{iwazaki3} with cross sections $(R'_a)^2$ less than $R_a^2$. 
Thus, the energies $\dot{W}(R'_a)^2/\lambda_a^2$ of the radiations are smaller than $10^{42}$erg/s,
although they are still consistent with the observed values $\sim 10^{40}$erg/s.  ) 
Therefore, the axion stars can pass through the geometrically thin accretion disk,
emitting large amount of the energies observed in the FRB 121102. Such emissions may arise each time
the axion star orbiting the black hole collides with the disk.


\section{characteristic features of repeating FRB}
\label{5}
\subsection{narrow bandwidth}
Now  we explain the features of the repeating FRB by using our model.
First, we explain that the bursts have finite bandwidths. The feature is common to the non repeating FRBs in our model.
The bursts are emitted by harmonically oscillating electrons with the frequency $\nu_{\rm int}=m_a/2\pi$.
 It apparently seems that the radiations are monochromatic.  
As we mentioned above, the electrons oscillate in one dimensional direction parallel to both the electric and magnetic field. 
Their transverse motions
are inhibited as long as the magnetic field is sufficiently strong for the cyclotron energy $eB/m_e$ to be much larger than thermal energies $\sim T$. 
Numerically, $eB/m_e\simeq 4\times10^3\mbox{eV}(eB/10^{11}\mbox{G})>T=10^2\mbox{eV}(T/10^6\mbox{K})$.
When the collision begins to takes place, the energies $p^2/2m_e\simeq (eE/m_a)^2/2m_e$ of 
the oscillating electrons are much larger than the initial temperature $T_{\rm ini}$ of electrons.
But, 
the temperature rapidly grows with time, because the energies of the oscillations are thermalized.
Eventually, the 
temperature approaches a critical one $T_c$ at which the coherent oscillations are disturbed   
and the coherent emissions are stopped.
The critical temperature is given by

\begin{equation}
 T_c=\frac{p^2}{2m_e}\simeq \frac{1}{2m_e}\Bigl(\frac{eE}{m_a}\Bigr)^2
\simeq 0.9\times 10^4\mbox{eV} (\sim 10^8\mbox{K} )\Big(\frac{3.5\times10^2\,\mbox{km}}{R_a}\Big)^4
\Bigl(\frac{0.6\times10^{-5}\mbox{eV}}{m_a}\Bigr)^4\Bigl(\frac{eB}{10^{11}\mbox{G}}\Bigr)^2
\end{equation}

That is, the critical temperature is equal to the oscillation energies of electrons. 
Thus, when we observe the radiations,
they have finite band widths owing to the thermal Doppler effects on the radiations.
In addition to the thermal Doppler effect,
the radiations are Doppler shifted
because the electron gases, which emit the radiations, rotate with the velocity $\vec{v}$ in the disk relative to the observer.
The frequencies $\nu_c$ ( called as center frequencies ) of the radiations Doppler shifted are given by 
$\nu_c=\nu_{\rm int}\sqrt{1-v^2}/(1-v\cos\theta)$ ( $\simeq \nu_{\rm int}(1+v\cos\theta) $ when $v\ll 1$ ) with
$\nu_{\rm int}=m_a/2\pi$ and $v=|\vec{v}|$.
$\theta$ denotes the angle between
the direction of the velocity $\vec{v}$ and the line of sight.
Therefore,
the spectrum $S(\nu)$ of the radiations is given by 

\begin{equation}
\label{S}
S(\nu)\propto \exp\Big(-\frac{(\nu-\nu_c)^2}{2(\delta\nu)^2}\Big)
\end{equation}
with the bandwidth $\delta\nu=\nu_c\sqrt{T_c/m_e}$. 
The formula can be applied to the radiations emitted by the accretion disk with the velocity $\vec{v}$
orbiting a black hole  or the radiations emitted by neutron stars moving with the velocity $\vec{v}$ relative to
the earth.

In the initial stage of the collision, the velocity $p/m_e$ of the oscillating electrons would be
much higher than thermal velocity $v_{\rm th}=\sqrt{2T_{\rm ini}/m_e}$. Thus, the thermal fluctuations do not disturb the oscillations.
But,
the temperature $T$ rapidly grows with time and 
the velocity $v_{\rm th}\propto \sqrt{T}$ also increases so that the bandwidth  $\delta\nu\propto \sqrt{T}$
increases.
Eventually, 
the coherent radiations ( FRBs ) stop at $T=T_c$ where the coherent emissions are thermally disturbed.
Therefore, it turns out that the bandwidths of the FRBs are determined by both the velocity $v$ of the accretion disk and
the critical temperature such as $\delta\nu=\nu_c\sqrt{T_c/m_e}$. For instance,
when the center frequency is equal to $\nu_c=3$GHz, the observed bandwidth is given by

\begin{equation}
\label{width}
\delta\nu=\nu_c\sqrt{\frac{T_c}{m_e}}\simeq 430\mbox{MHz}\frac{eB}{10^{11}G}.
\end{equation} 

The recent observation\cite{finite} of the FRB 121102 shows that spectra of several bursts are characterized by the Gaussian form
just like eq(\ref{S}) and shows that
the bandwidths are roughly given by $500$MHZ when
the center frequency $\nu_c=3$GHz. 
( A variety of the bandwidths of the bursts with almost identical $\nu_c$ could be explained by a variety of the magnetic fields
because $\delta\nu\propto \nu_c B$. )
Similarly, it seems that the spectra of the FRBs No.$13$ and No.$16$ in the ref.\cite{2GHz} show 
the narrower bandwidth $\sim 300$MHz for lower center frequency $\sim 2$GHz. That is, the bandwidth is proportional to the center frequency as shown in	 
the formula in eq(\ref{width}).
The formula can be also applied to the bursts FRB 121102 observed with lower center frequencies $\nu_c\simeq 1.2$GHz.
Although their bandwidths have not been precisely measured, the spectra of the FRBs No.$9$ and No.$10$ in the ref.\cite{rep}
show that their bandwidths are given by $\sim 200$MHz or less for the radiations with the center frequency $\sim 1.2$GHz.
Therefore, we find that 
the observed bandwidths of the repeating FRBs 121102 can be roughly explained by our formula $\delta\nu=\nu_c\sqrt{T_c/m_e}$
if we take the magnetic fields such as $eB\sim 10^{11}$G for these bursts.
Namely, it seems that the observed bandwidths $\delta\nu$ are proportional to $\nu_c$,

\begin{eqnarray}
\delta\nu&\sim& 500\mbox{MHz} \quad \mbox{for center frequencies}\quad \nu_c\sim 3\mbox{GHz in the ref.\cite{finite}} \nonumber \\
\delta\nu&\sim& 300\mbox{MHz} \quad \mbox{for center frequencies}\quad \nu_c\sim 2\mbox{GHz in the ref.\cite{2GHz}} \nonumber \\
\delta\nu&\sim& 200\mbox{MHz} \quad \mbox{for center frequencies}\quad \nu_c\sim 1.2\mbox{GHz in the ref.\cite{rep}}.
\end{eqnarray}

\vspace{0.1cm}
We have shown in our model that both spectra of the repeating and non repeating FRBs have finite bandwidths owing to the thermal Doppler effect.
The observed bandwidths are wide such as $200\rm MHz\sim 500$MHz. This is caused by the high critical temperatures of the electron gases
e.g.  $T_c\sim 10^{8}$K. Such high temperatures are necessary to disturb the coherent oscillations of the electrons
supported by strong magnetic fields $eB\sim 10^{11}$G. 
It is important to remember that the production of the large amount of the energies $\sim 10^{42}$ erg/s in FRBs is caused by
similarly strong magnetic fields $\sim 10^{11}$G. 
Therefore, both of the narrow bandwidths and the large amount of the energies
originate in the strong magnetic fields of the order of $10^{10}$G present in the accretion disk.

 
There is a variety of the center frequencies of the bursts in FRB 121102.
The varieties arise from a variety of local velocities with which accretion disks rotate.
Since the observed center frequencies $\nu_c$ are given by
 $\nu_c=\nu_{\rm int}\sqrt{1-v^2}/(1-v\cos\theta)$,
the presences of the bursts with the large center frequencies $\sim 3$GHz compared with the intrinsic one $\nu_{\rm int}\sim 1.4$GHz 
implies that the velocities $v$ of the accretion disk emitting the bursts must be larger than $v=0.6$
and that the angle $\theta$ is small such as $\theta< 0.2$. 
We speculate that there are few bursts with much higher center frequencies than $3$GHz, because
the angle $\theta$ must be fairly small for such bursts.
Actually, there have been no bursts observed with frequencies higher than $4$GHz. 
( After submitting the paper, we found a report\cite{atel} on the observation of $15$ bursts with the frequencies, $4\rm GHz\sim 8GHz$ in FRB 121102. 
All bursts show narrow bandwidths, a hundred MHz $\sim 1$GHz. Although the detail analysis of the observation has not been published, it seems that
the narrow bandwidths are consistent with our prediction $\delta\nu\sim 1\rm GHz$ for $\nu_c=6$GHz. 
These bursts with high center frequencies $\sim 6$GHz
arise from the locations in the accretion disk whose velocities are very large $v \geq 0.95$. )

\vspace{0.1cm}
The collisions generating the repeating FRBs with lower center frequencies $< 100$MHz take place in the disks
with the velocities larger than $0.99$ and $\theta\simeq \pi$. Thus, such collisions would be very rare. Actually, there have been no observations of
the bursts with such low frequencies.   

We would like to mention that in our model, 
non repeating FRBs arises from the collisions between axion stars and neutron stars.
In the case, Doppler effects owing to the velocities $v$ ( $\ll 1$ ) of the electron gas are small.
Thus, it is expected that there are no such a variety of center frequencies.
Actually, the frequencies of non repeating FRBs are limited in a range $700\rm MHz \sim 1.6 GHz$.
No non repeating bursts have been observed with much higher ( or lower frequencies ) such as $5$GHz ( or $200$MHz ).
Additionally, neutron stars in general possess stronger magnetic fields $\sim 10^{12}$G than the 
magnetic fields $\sim 10^{11}$G in the accretion disk used in the above estimations.
It implies that emission rates $\dot{W}$ of non repeating FRBs are large and their bandwidths $\delta \nu$ are wide
compared with those of the repeating FRB 121102.
Actually, the finiteness of the bandwidths in non repeating FRBs have not yet been observed.
Probably, their bandwidths may be wider than $300$MHz.
Furthermore,  
the observed volatility of the spectral indices of non repeating FRBs and the sharp cutoff in the spectrum of the FRB 170107 indicate the finiteness
of the bandwidths. 
All of the observations are consistent with our production mechanism of the FRBs.

\subsection{interval of repeating bursts}
One of the characteristic features of the repeating bursts is that there are no regularities in the time interval of the bursts.
But, there is a tendency that the bursts are clustered. Several bursts continuously happen in a short period and after that there is a long period, 
in which no bursts take place.  Then, a cluster of several bursts again take place.
We will explain these features in the FRB 121102.
We suppose that there would be many dark matter axion stars condensing in an early type galaxy like the one associated with the FRB 121102.
Some of them would fall into the black hole. 
A falling axion star collides the accretion disk several times,
orbiting the black hole. After the several collisions, the axion star evaporates or is absorbed in the black hole. 
Then, next axion star collides the disk in a similar way.
Each collision generates a burst.
Thus, the emissions of the repeating FRBs are clustered. After the clustered emissions, there is a long or short period without
no bursts. The length of the period depends on when the next axion star collides the disk.

The sequence of the emissions clustered may depends on how collisions take place. That is, the sequence depends on 
whether or not the orbit plane of the axion star changes after its collision with the disk. 
If the orbit plane of
the axion star changes after the collisions, the sequence of the collisions may be not regular.
On the other hand, if the orbit plane does not change after each collision, the sequence may be regular. But,
even if the sequence is regular, observed sequence is not necessarily regular because 
the radiations are possibly intercepted by the black hole. Whether or not the intercept take places depends on the
geometrical relation between the location of
a collision and that of the black hole. 
In this way, whether or not the regularities in the sequence of the bursts can be seen depends
on how the collisions take place.

In the ideal case that the orbit plane is located almost perpendicular to the line of sight, the regular sequence of the emissions
can be observed. In the case the time interval between an emission and a subsequent emission may be shorter than 
the preceding time interval. 
This is because the axion star orbits and falls into the black hole 
accelerating its speed. That is, the rotation period becomes shorter 
each time the axion star collides the disk.

It is very intriguing that the recent report\cite{atel} shows such a regularity in the time intervals.
The time intervals between the adjacent emissions of the bursts
$11\rm B, 11C, 11D$ and  $11\rm E$ ( we used the notation in the ref\cite{atel} ) are $38\rm sec., 21sec.$ and $11\rm sec.$, respectively.
Furthermore, the time intervals between the adjacent emissions of the bursts $11\rm G, 11 H, 12B$ and $11\rm I$ are $94\rm sec., 51 sec.$ and $27 \rm sec.$, respectively.
The observation strongly supports our model for the production mechanism of the repeating FRBs.

\section{discussion and summary}
\label{6}
Finally we would like to discuss how the observed frequencies $\nu_{\rm ob}$ are related to the intrinsic frequency $\nu_{\rm int}=m_a/2\pi$. 
When the axion stars oscillating with the frequency $\nu_{\rm int}=m_a/2\pi$ collide neutron stars or black holes, 
their velocities $v_0$ get near the light velocity. Thus, the electron gases in the neutron stars or accretion disks 
feel electric fields oscillating with higher frequencies $\nu'=\nu_{\rm int}\gamma$ ( $\gamma=1/\sqrt{1-v_0^2} >1$ ) than the intrinsic one $\nu_{\rm int}$. Then,
the electrons oscillate with the frequency $\nu'$ and emit the radiations with the frequency $\nu'$.
On the other hand, when we observe the radiations, the radiations  
are gravitationally red shifted ( $\nu''=\sqrt{g_{t,t}}\nu'$ with $g_{t,t}=\sqrt{1-2GM/r_0}<1$ where $r_0$ denotes the distance
of the emission location from the center of neutron star or black hole, ). This is because we are free from the gravities of the neutron stars or black holes with mass $M$. 
Furthermore, 
when the electron gases move with the velocity $v$ relative to us, the radiations from the gases are blue or red shifted by the factor $\sqrt{1-v^2}/(1+v\cos\theta)$.
Additionally the radiations are cosmologically red shifted; $\nu\to \nu/(1+z)$. 
Consequently, the observed frequency $\nu_{\rm ob}$ is given by $\nu_{\rm ob}=(1+z)^{-1}\sqrt{g_{t,t}}\nu_{\rm int}\gamma\sqrt{1-v^2}/(1+v\cos\theta)$. 
Here we point out that  in general $\sqrt{g_{tt}}\gamma\simeq 1$.
This is because $v_0\simeq \sqrt{2GM/r_0}$. 
Therefore, we may obtain $\nu_{\rm ob}\simeq (1+z)^{-1}\sqrt{1-v^2}/(1+v\cos\theta)\nu_{\rm int}$.
Non repeating FRBs arise in the collisions between the axion stars and neutron star. 
In the case the velocity $v$ of the neutron star's atmospheres is much small, i.e. $v\ll 1$.
Then, we obtain $\nu_{\rm ob}=\nu_{\rm int}/(1+z)$.
Thus, for example, we can understand that the non repeating FRB 110523 observed\cite{700} with the low frequencies $\sim 700$MHz 
arises at far-distant location e.g. $z\sim 0.5$.  
Probably FRBs at much far distances, e.g. $z>2$ would be not observed since they are very faint. 
Thus, we can predict that there are no observations of non repeating FRBs with frequencies higher than $5$GHz or lower than a few hundred MHz.

\vspace{0.2cm}   
In summary, we have proposed a model for the production mechanism of the repeating FRB 121102; the bursts are emitted by the collisions between axion stars
and geometrically thin magnetized accretion disk around a black hole in the dwarf galaxy observed.
The collisions take place several times because the axion star orbits the black hole.
The radiations emitted in the collisions are linearly polarized dipole radiations. They arise from 
the coherent harmonic oscillations of electrons with the frequency $\nu_{\rm int}=m_a/2\pi$. The oscillations are induced 
by the oscillating electric fields generated in the axion stars
under strong magnetic fields $\sim 10^{11}$G. The observed large amount of their energies can be explained by
such strong magnetic fields. On the other hand, the emissions of the coherent radiations are terminated by the thermal fluctuations because
the thermalization of the oscillation energies makes the temperature of electron gas increase.
We have shown that
the observed narrow bandwidths $\delta\nu$ are caused by the thermal Doppler effects on the radiations emitted by
the harmonically oscillating electrons.
Hence, the bandwidths are proportional to the center frequency $\nu_c$ of the radiations.   
It is notable that the observed bandwidths $200\rm MHz\sim 500$MHz 
can be well explained by the strong magnetic fields $\sim 10^{11}$G stated above.
Similarly non repeating FRBs are expected to show
the spectral forms $S(\nu)\propto \exp\big(-(\nu-\nu_c)^2/2(\delta\nu)^2\big)$.
In our model the center frequencies $\nu_c$ are given by the axion mass $m_a$ such that $\nu_c=\frac{m_a}{2\pi(1+z)}$
in the non repeating FRBs. Thus,
the determination of red-shift $z$ of the sources in non repeating FRBs can lead to the determination of the axion mass.

In this way, almost of all features of the repeating FRB 121102 can be well understood both qualitatively and quantitatively 
in our model. 
The recent observation\cite{rotation} showing the presence
of highly magnetized gases around the source of FRB 121102 supports our model.

.

\vspace{0.2cm}
The author
expresses thanks to Prof. M. Kawasaki, University of Tokyo, Prof. F. Takahashi, University of 
Tohoku and Prof. T. Terasawa, Riken for useful comments and also expresses thanks to
members of KEK for their hospitality.




\begin{thebibliography}{99}
\bibitem{frb}D. R. Lorimer, M. Bailes, M. A. McLaughlin, D. J. Narkevic, F. Crawford, 
Science, 318 (2007) 777.  \\
E. F. Keane, D. F. Ludovici, R. P. Eatough, et al.,
MNRAS, 401 (2010) 1057.
\bibitem{frb2}L. G. Spitler, J. M. Cordes, J.W.T. Hessels, et al., ApJ, 790 (2014) 101.
\bibitem{rep}L.G. Spitler, P. Scholz, J.W.T. Hessels, et al. Nature,  531 (2016) 202.
\bibitem{2GHz}P. Scholz, L.G. Spitler, J.W.T. Hessels, et al. arXiv:1603.08880.
\bibitem{re3G}S. Chatterjee, et al. Nature, 541 (2017) 58.
\bibitem{re1.6G}B. Morcote, et al. arXiv:1701.01099.
\bibitem{op}S.P. Tendulkar, et al. arXiv:1701.01100.
\bibitem{finite}C.J. Low, et al. arXiv:1705.07553.
\bibitem{170107}K.W. Bannister, et al. arXive:1705.07581.
\bibitem{model}T. Totani, PASJ, L21 (2013) 65.\\
K. Kashiyama, K. Ioka, and P. Meszaros, ApJL, L39 (2013) 776.\\
S. B. Popov and K. A. Postnov arXiv:1307.4924.\\
H. Falcke and L. Rezzolla, A and A, A137 (2014) 562.\\
A. Loeb, Y. Shvartzvald and D. Maoz, MNRAS, L46 (2014) 439.\\
K. W. Bannister and G. J. Madsen, MNRAS, 353 (2014) 440. \\
B.D. Metzger, E. Berger and B. Margalit, arXve:1701.02370.
\bibitem{iwazaki}A. Iwazaki, Phys.Rev. D91 (2015) no.2, 023008.
\bibitem{t}I. I. Tkachev, JETP Lett. 101 (2015) no.1, 1. \\ S. Raby, Phys. Rev. D.94 (2016) 103004.
\bibitem{accretion}M.C. Begelman and J. Silk,  Mon. Not. Roy. Astron. Soc. 464 (2017) no.2, 2311.
\bibitem{axion}R. D. Peccei and H. R. Quinn, Phys. Rev. Lett. 38 (1977) 1440.\\
S. Weinberg, Phys. Rev. Lett. 40 (1978) 223.\\
F. Wilczek, Phys. Rev. Lett. 40 (1978) 279.	
\bibitem{rotation}D. Michilli, et al. Nature 553. (2018) 182.
\bibitem{axion2}E. Seidel and W.M. Suen, Phys. Rev. Lett. 72 (1994) 2516.
\bibitem{osci}M. Alcubierre, R. Becerril, F. S. Guzman, T. Matos, D. Nu�nez and L. A. Ure�na-L�opez,
Class.Quant.Grav. 20 (2003) 2883.\\
G. Fodor,  P. Forgacs and M. Mezei, Phys. Rev. D82 (2010) 044043.
\bibitem{bs}S. Liebling and C. Palenzeula,  Living Rev. Rel. 15 (2012) 6. 
E. Braaten, A. Mohapatra and H. Thang,  Phys. Rev. Lett. 117 (2016)  121801. 
R. Ruffini and S. Bonazzola, Phys. Rev. 187  (1969) 1767. 
\bibitem{kolb}E. W. Kolb and I. I. Tkachev, Phys. Rev. Lett. 71 (1993) 3051;
Astrophys. J. 460 (1996) L25.
\bibitem{axions}P. Jetzer, Phys. Rep. 220 (1992).\\
E. Seidel and W.M. Suen, Phys. Rev. Lett. 66 (1991) 1659.\\
A. Iwazaki, Phys. Lett. B451 (1999) 123.\\
F. S. Guzman and L. A. Urena-Lopez, Astrophys. J. 645 (2006) 814.
\bibitem{iwa}A. Iwazaki, Phys. Lett. B489 (2000) 353.\\
J. Barranco and A. Bernal, Phys. Rev. D83 (2011) 043525.
\bibitem{iwazaki2}A. Iwazaki, arXiv:1412.7825.
\bibitem{sol}F. Kling and A. Rajaraman, arXiv:1706.04272.
\bibitem{iwazaki3}A. Iwazaki, arXiv:1512.06245.
\bibitem{700}K. Masui. et al. Nature 528 (2015) 525.
\bibitem{atel}V. Gajjar, et.al. ATel No.10675.
\end{thebibliography}
\end{document}